\documentclass[preprint,showpacs,showkeys,preprintnumbers,amsmath,amssymb]{revtex4}

\usepackage{graphicx}
\usepackage{dcolumn}
\usepackage{bm}
\newcommand{\beq}{\begin{equation}}
\newcommand{\eeq}{\end{equation}}
\newcommand{\beqa}{\begin{eqnarray}}
\newcommand{\eeqa}{\end{eqnarray}}
\newcommand{\lam}{\lambda}

\newcommand{\rh}{\rho}
\newcommand{\ga}{\gamma}

\newcommand{\al}{\alpha}

\newcommand{\om}{\omega}

\newcommand{\la}{\langle}
\newcommand{\ra}{\rangle}

\begin{document}

\title{Non-Markovian Quantum Trajectories Versus Master Equations: Finite Temperature Heat Bath}

\author{Ting Yu\footnote{Also at Department of Physics,
Queen Mary College, University of London, London, E1 4NS,
UK}}
\email{ting@pas.rochester.edu} \affiliation{ Rochester Theory
Center for Optical Science and Engineering \\and\\ Department of
Physics and Astronomy, University of Rochester, Rochester, NY
14627, USA\\}

%
\date{29 March 2004}

\begin{abstract}

The interrelationship between the non-Markovian stochastic
Schr\"odinger equations and the corresponding non-Markovian master
equations is investigated in the finite temperature regimes. We
show that the general finite temperature non-Markovian
trajectories can be used to derive the corresponding non-Markovian
master equations. A simple, yet important solvable example is the
well-known damped harmonic oscillator model in which a harmonic
oscillator is coupled to a finite temperature reservoir in the
rotating wave approximation. The exact convolutionless master
equation for the damped harmonic oscillator is obtained by
averaging the quantum trajectories relying upon no assumption of
coupling strength or time scale. The master equation derived in
this way automatically preserves the positivity, Hermiticity and
unity.

\end{abstract}

\pacs{03.65.Yz, 42.50.Lc, 05.40.Jc}

\keywords{Open quantum systems; Non-Markovian quantum
trajectories, Master equations}
\maketitle

\section{Introduction}\label{introduction}
Open systems are generic models for the study of quantum dynamics
in the sense that a real quantum system is either difficult to be
isolated from the influence of its environment or is deliberately
put in touch with some purposely engineered devices in order to
make a measurement \cite{GAR,CAR1}. In any case, the system and
the environment, initially independent, will become entangled due
to the interaction, then the system state will not remain in a
pure state and the dynamics is described by a non-unitary process.
In many physically relevant cases, a typical open quantum system
normally involves a small system of interest coupled to a large
system with large number of degrees of freedom, commonly known as
heat bath, reservoir or generally environment. Traditionally, the
dynamics of a system of interest is described by a Lindblad master
equation which can often be derived if the standard Born-Markov
approximation is assumed \cite{CAR1,COH,LIN,GKS}. Moreover, it
turns out that such a Lindblad master equation also allows to be
unravelled into various stochastic Schr{\"o}dinger equations known
as quantum trajectories \cite{CAR,DM,GP,GZ,WM,PK,PER}. Both the
master equations of the Lindblad form and their stochastic
unravelings have become an integral part of theories of open
quantum systems. However, when the heat bath memory effects are
relevant, such as in the cases of high-Q cavity, atom laser or
structured environment, where the Born-Markov approximation ceases
to be valid, hence the dynamics of the open quantum system must be
described by a non-Markovian process \cite{NM1,NM2}. It is long
known that the derivation of a non-Markovian master equation is a
formidable task \cite{NAK,ZWA}.


Recent research on open quantum systems has suggested that,
alternatively, the non-Markovian dynamics may be described by a
diffusive stochastic Schr\"odinger equation known as non-Markovian
quantum trajectories or quantum state diffusion equation
\cite{DS,DGS,DGS1,YDGS,YDGS2,BKP,JCW,CRE,BU,GW,BA,YDGS1,SY}. For
an $N$-dimensional Hilbert space, the stochastic Schr{\"o}dinger
equation evolves an $N$-dimensional vector, so offers numerical
advantages over a master equation which evolves an $N\times N$
density matrix. Very recently, we have shown that, beyond the
numerical advantages and the conceptual merits, the non-Markovian
stochastic Shr{\"o}dinger equations may also provide a powerful
tool for deriving the corresponding non-Markovian master
equations. This idea has been explored in several distinct cases
including a two-level atom interacting with a zero-temperature
heat bath \cite{YDGS,YDGS1}, and a Brownian particle coupled
linearly to a finite temperature heat bath via the position
variable commonly known as quantum Brownian motion model
\cite{SY,YDGS2,HR}. The purpose of this paper is to extend this
research to more general finite temperature regimes where the
Lindblad operator is not a Hermitian operator. In this paper, we
take the damped harmonic oscillator as our primary example. This
simple model is of great interest in quantum optics because it is
an essential ingredient in the theoretical investigations of
various quantum optical experiments. We show here in detail that
the non-Markovian quantum trajectories allow the derivation of the
exact convolutionless master equation irrespective of the coupling
strength, the separated time scales, or the special distributions
of the environmental frequencies.

The organization of the paper is as follows. In Sec. ~\ref{sto},
we introduce both zero and finite temperature stochastic
Schr\"odinger equations. In Sec. \ref{master}, we show how to
derive an exact master equation from the finite temperature
quantum trajectories. In Sec. \ref{dmaped}, we establish the
stochastic Schr\"odinger equation for the damped harmonic
oscillator and show that the corresponding exact convolutionless
master equation can be obtained by averaging the solutions to the
non-Markovian stochastic Schr{\"o}dinger equation without any
approximations, in particular, without Markov approximation. We
conclude this paper in Sec.~\ref{sect5}
\section{Non-Markovian stochastic Schr\"odinger equation}
\label{sto}
\subsection{Zero-temperature ($T=0$)}
\label{zero}
Our model consists of a system of interest coupled linearly to a
large number of harmonic oscillators with distributed
eigenfrequencies $\omega_{\lambda}$ and creation and annihilation
operators $b^{\dagger}_{\lambda}, b_{\lambda}$. The quantum
Hamiltonian for the system plus reservoir can be typically written
as (we set $\hbar=1$)
\beqa\label{totalH} H_{\rm tot}&=& H_{\rm sys} + H_{\rm int} +
H_{\rm bath}\nonumber \\&=& H_{\rm sys} +\sum_\lambda (g^*_\lambda
L^\dag b_\lambda +g_\lam L b^\dag_\lambda)+ \sum_\lambda
\omega_\lambda b^\dag_\lambda b_\lambda, \eeqa where $g_\lambda$
are the coupling constants and the system operator $L$ coupled to
the environment is often called the Lindblad operator.

For the open quantum system described by (\ref{totalH}), the
linear non-Markovian quantum state diffusion (QSD) equation has
been derived from the formal solution of Schr\"odinger equation
for the total system in a special representation \cite{DGS}:
\begin{equation}
\label{schrequ} i\partial_t|\Psi_t\rangle =H_{\rm
tot}(t)|\Psi_t\rangle,
\end{equation}
where $|\Psi_t\rangle$ stands for the pure state vector for the
total system and $H_{\rm tot}(t)$ is the Hamiltonian $H_{\rm tot}$
in the interaction representation with respect to the free bath
Hamiltonian: \beqa H_{\rm tot}(t)&=&e^{iH_{\rm bath}t}\big( H_{\rm
sys} + \sum_\lambda (g^*_\lambda L^\dagger a_\lambda +
g_\lambda L a_\lambda^\dagger)\big) e^{-iH_{\rm bath}t}\nonumber\\
&=&H_{\rm sys} + \sum_\lambda \left (g^*_\lambda L^\dagger
a_\lambda e^{-i\omega_\lambda t}+ g_\lambda L a_\lambda^\dagger
e^{i\omega_\lambda t}\right). \eeqa In this subsection, let us
assume that the initial pure state of the system and the
environment is taken to be
\begin{equation}
|\Psi_0\rangle=|\psi_0\rangle\otimes |0_1\rangle\otimes
|0_2\rangle\cdots\otimes|0_\lambda\ra\otimes\cdots,
\end{equation}
with an arbitrary system state $|\psi_0\rangle$ and the
environment in the ground state $|0\rangle$. By using a Bargmann
coherent state basis for the environmental degrees of freedom:
$|z_{\lambda}\rangle=\exp\{z_{\lambda}a^{\dag}_{\lambda}\}|0_\lambda\rangle$
and the resolution of the identity
\begin{equation}
I_{\lambda}=\int \frac{d^2 z_{\lambda}}{\pi}{\rm
e}^{-|z_{\lambda}|^2} |z_{\lambda}\rangle\langle z_{\lambda}|,
\end{equation}
the total state  $|\Psi_t\rangle$ can be expressed  as
\begin{equation}
|\Psi_t\rangle=\int \frac{d^2 z}{\pi}{\rm
e}^{-|z|^2}|\psi_t(z^*)\rangle\otimes|z\rangle,
\end{equation}
where $|z\rangle=|z_1\rangle\otimes z_2\rangle\otimes \cdots
\otimes|z_{\lambda}\rangle \cdots$, $d^2z=d^2z_1d^2z_2\cdots
d^2z_\lambda\cdots$ and $|z|^2=\sum_\lambda |z_\lambda|^2$. Then
the resultant pure state for the system of interest
$|\psi_t(z^*)\rangle$ was shown to satisfy the following equation
\cite{DS,DGS}:
\begin{equation}\label{NMQSD}
\partial_t \psi_t = -i H_{\rm sys} \psi_t + Lz^*_t \psi_t -L^\dagger
\int_0^t ds\;\alpha(t-s) \frac{\delta\psi_t}{\delta z^*_s}.
\end{equation}
where $\alpha(t-s) =
 \sum_\lambda |g_\lambda|^2 e^{-i\omega_\lambda(t-s)}$ is
the bath correlation function, and  $z_t^* = -i\sum_\lambda
g_\lambda^* z_\lambda^* e^{i\omega_\lambda t}$ is a colored,
complex Gaussian noise with ${\mathcal M}[z_t]={\mathcal M}[z_t
z_s]=0$ and ${\mathcal M}[z_t^* z_s] = \alpha(t-s)$. Note here
${\mathcal M}[\,\cdot\,]$ is the statistical mean over the
Gaussian process $z_t$. By construction, \beq \rho_t = {\mathcal
M}[|\psi_t\ra\la \psi_t|]=\int
\frac{d^2z}{\pi}e^{-|z|^2}|\psi_t\ra \la \psi_t|.\eeq

The linear non-Markovian QSD equation (\ref{NMQSD}) is valid for a
zero-temperature heat bath, and also for a finite temperature heat
bath with the condition $L=L^\dagger$. In general, if $L\neq
L^\dag$, we shall see below that the finite temperature
non-Markovian QSD equation takes a more complicated form.

\subsection{Finite temperature ($T\neq 0$)}
\label{nonzero}
The finite temperature stochastic Schr{\"o}dinger equation or
non-Markovian QSD equation can be obtained by mapping the system
with the total Hamiltonian (\ref{totalH}) and an initial thermal
state to an extended system with a vacuum state such that the zero
temperature stochastic Schr{\"o}dinger equation for the extended
system is equivalent to the situation where the heat bath is at
finite temperature \cite{DGS,KK}. To be specific, for the total
Hamiltonian (\ref{totalH}), we assume that the heat bath is in a
thermal equilibrium state at temperature $T$, with density
operator \beq \rho_{\rm bath}(0)=\frac{e^{-\beta H_{\rm
bath}}}{Z}, \eeq where $Z={\rm Tr}[e^{-\beta H_{\rm bath}}]$ is
the partition function and $\beta=1/k_BT$.

Now we introduce a fictitious heat bath with the bosonic operators
$c_\lam, c_\lam^\dag$, which has no direct interaction with the
system, ensuring that after tracing over the fictitious variables,
the initial state of the original bath will be in the thermal
state $\rh_{\rm bath}(0)$.  The total Hamiltonian with two
independent heat baths is given by \beq\label{totalH20} {\mathcal
H}_{\rm tot}=H_{\rm sys} +\sum_\lambda (g^*_\lambda L^\dag
b_\lambda +g_\lam Lb^\dag_\lambda)+ \sum_\lambda \omega_\lambda
b^\dag_\lambda b_\lambda -\sum_\lambda \omega_\lambda
c^\dag_\lambda c_\lambda. \eeq

The boson Bogoliubov transformation formally couples the system of
interest to two set of bosonic operators $d_\lam, d_\lam^\dag$ and
$e_\lam,e_\lam^\dag$ : \beqa
{b}_\lam&=&\sqrt{\bar{n}_\lam+1}{d}_\lam+\sqrt{\bar{n}_\lam}
{e}^\dag_\lam ,\\
{c}_\lam&=&\sqrt{\bar{n}_\lam+1}{e}_\lam+\sqrt{\bar{n}_\lam}
{d}^\dag_\lam ,\eeqa where $\bar{n}_\lam$ is the mean thermal
occupation number of quanta in mode $\om_\lam$: \beq
\bar{n}_\lam=\frac{1}{\exp(\hbar \omega_\lam/k_BT)-1}. \eeq

The transformed Hamiltonian ${\mathcal H'}_{\rm tot}$ in terms of
$d_\lam, e_\lam$ is then given by \beqa\label{totalH2} {\mathcal
H'}_{\rm tot}&=&H_{\rm sys} +\sum_\lambda
\sqrt{\bar{n}_\lam+1}(g_\lam^* L^\dag d_\lambda +g_\lam
Ld^\dag_\lambda)+ \sum_\lambda \omega_\lambda d^\dag_\lambda
d_\lambda\nonumber\\&& +\sum_\lambda \sqrt{\bar{n}_\lam}
(g_\lam^*L^\dag e^\dag_\lambda +g_\lam Le_\lambda)-\sum_\lambda
\omega_\lambda e^\dag_\lambda e_\lam . \eeqa Thus we have mapped
the finite temperature problem into a zero temperature one  with
the initial vacuum state denoted by
$|0\ra=|{0}\ra_d\otimes|{0}\ra_e$, satisfying $d_\lam|{0}\ra=0,
e_\lam|{0}\ra=0$. With the new Hamiltonian (\ref{totalH2}), the
finite temperature problem has been reduced to a zero temperature
one. Thus the resultant pure state for the system of interest
$\psi_t=|\psi_t(z^*,w^*)\rangle$ satisfies the following
stochastic Schr{\"o}dinger equation with two independent noises
$z^*_t,w^*_t$: \beqa\label{NMQSD1}
\partial_t \psi_t &=& -i H_{\rm sys} \psi_t + Lz^*_t \psi_t -L^\dag\int_0^t ds\;\alpha_1(t-s)
\frac{\delta\psi_t}{\delta z^*_s}\nonumber\\
&&+L^\dag w^*_t \psi_t -L\int_0^t ds\;\alpha_2(t-s)
\frac{\delta\psi_t}{\delta w^*_s}, \eeqa where
\beqa \alpha_1(t-s)
&=& \sum_\lambda (\bar{n}_\lambda+1)|g_\lambda|^2
e^{-i\omega_\lambda(t-s)},\\
\alpha_2(t-s) &=& \sum_\lambda \bar{n}_\lambda |g_\lambda|^2
e^{i\omega_\lambda(t-s)}\eeqa are the bath correlation functions
and \beqa
 z_t^*&=&-i\sum_\lam\sqrt{\bar{n}_\lam+1}\,\,g^*_\lam z_\lam^*
e^{i\om_\lam t},\\
w_t^*&=&-i\sum_\lam\sqrt{\bar{n}_\lam}\,\,g^*_\lam w_\lam^*
e^{-i\om_\lam t} \eeqa are two independent, colored, complex
Gaussian noises satisfying \beqa {\mathcal M}[z_t]&=&{\mathcal
M}[z_t z_s]=0,\,\,{\mathcal M}[z_t^* z_s] = \alpha_1(t-s),\\
{\mathcal M}[w_t]&=&{\mathcal M}[w_t w_s]=0,\,\,{\mathcal
M}[w_t^*w_s] = \alpha_2(t-s). \eeqa
 Note here ${\mathcal M}[\,\cdot\,]$
is the statistical mean over the Gaussian processes $z^*_t$ and
$w^*_t$. Again, by construction, $\rho_t = {\mathcal
M}[|\psi_t\ra\la \psi_t|]$. In the zero temperature limit
$T\rightarrow 0$, we have $\alpha_1(t-s)\rightarrow \sum_\lambda
|g_\lambda|^2\exp(-i\om_\lam(t-s))$ and $\alpha_2(t-s)\rightarrow
0$, then equation (\ref{NMQSD1}) reduces to the simple
zero-temperature case (\ref{NMQSD}). However, in this paper,
without explicit statement, we will always work in the finite
temperature regimes. From equation (\ref{NMQSD1}), we see that the
finite temperature heat bath has induced both the spontaneous
transitions and stimulated transitions, moreover, it also gives
rise to an absorptive process caused by taking thermal quanta from
the heat bath.

The stochastic Schr\"{o}dinger equation (\ref{NMQSD1}) can be
greatly simplified by using the ansatz, \beqa\label{op}
\frac{\delta\psi_t}{\delta
z^*_s}&=&O_1(t,s,z^*,w^*)\psi_t,\\
\frac{\delta\psi_t}{\delta w^*_s}&=&O_2(t,s,z^*,w^*)\psi_t.
\label{op1}\eeqa Then equation (\ref{NMQSD1}) takes a more compact
form: \beqa \label{nmcom}
\partial_t \psi_t &=& -i H_{\rm sys} \psi_t +
Lz^*_t \psi_t -L^\dag \bar{O}_1(t,z^*,w^*)\psi_t\nonumber\\
&&+L^\dag w^*_t \psi_t -L \bar{O}_2(t,z^*,w^*)\psi_t,\eeqa where
$\bar{O}_i\,\,(i=1,2)$ denote \beq
\label{obar12}\bar{O}_i(t,z^*,w^*)=
\int^t_0\alpha_i(t-s)O_i(t,s,z^*,w^*)ds,\,\,\,(i=1,2).\eeq

We can determine the operators $O_{1,2}(t,s,z^*,w^*)$ in
(\ref{op}) and (\ref{op1}) from the following `consistency
conditions',
\begin{equation}\label{consistency}
\partial_t \frac{\delta\psi_t}{\delta z^*_s} =
\frac{\delta}{\delta z^*_s} \partial_t\psi_t,  \,\,\,\partial_t
\frac{\delta\psi_t}{\delta w^*_s} = \frac{\delta}{\delta w^*_s}
\partial_t\psi_t,
\end{equation}
together with the initial conditions:
\begin{equation}
\label{initialcond} O_1(t=s,s,w^*,z^*) = L,\,\,\,
O_2(t=s,s,w^*,z^*) =L^\dag.
\end{equation}

From the consistency conditions, we may get the evolution
equations for $O_1=O_1(t,s,z^*,w^*)$ and $O_2=O_2(t,s,z^*,w^*)$
\beqa\label{O11}
\partial_t O_1&=& \left[-iH_{\rm sys} +
Lz_t^* +L^\dag w_t^*- L^\dagger{\bar O}_1 -L\bar{O}_2,
O_1\right]\nonumber\\
&& - L^\dagger \frac{\delta {\bar O}_1}{\delta
z^*_s}-L\frac{\delta {\bar O}_2}{\delta z^*_s}, \eeqa and
\beqa\label{O12}
\partial_t O_2&=& \left[-iH_{\rm sys}
+ Lz_t^* +L^\dag w_t^*- L^\dagger{\bar O}_1- L\bar{O}_2,
O_2\right]\nonumber\\
&& - L^\dagger \frac{\delta {\bar O}_1}{\delta
w^*_s}-L\frac{\delta {\bar O}_2}{\delta w^*_s}. \eeqa

It should be remarked here that the O-operators can be determined
in many interesting situations \cite{DGS,SY}, and moreover, that
the approximate O-operators can always be obtained by invoking a
perturbation technique  \cite{YDGS,YDGS1,YDGS2}

\section{Non-Markovian master equation at finite temperature}
\label{master}
From the non-Markovian QSD equation (\ref{nmcom}), one may take
the statistical mean to derive the master equation, and gets
\beqa\label{generalmaster}
\partial_t \rho_t &=& -i[H_{\rm sys},\rho_t]
+[L,\,\, {\mathcal M}\{P_t {\bar O_1}^\dagger(t, z^*,w^*)\}]
-[L^\dagger,\,\,{\mathcal M}\{{\bar
O}_1(t,z^*,w^*) P_t\}]\nonumber\\
&&+[L^\dag,\,\,{\mathcal M}\{P_t{\bar O}_2^\dag(t,z^*,w^*)\}] -
[L,\,\,{\mathcal M}\{\bar{O}_2(t,z^*,w^*)P_t \}]. \eeqa This is
the general master equation at finite temperature. Although this
last result is still not a closed evolution equation for $\rho_t$,
it nevertheless shows how a convolutionless master equation may
result from our knowledge of the operators $O_i(t,s,z^*,w^*),\,\,
(i=1,2)$.

As we mentioned before, in the case of finite temperature with
$L\neq L^\dag$, the O-operators will generally contain noises
$z_t^*,w^*_t$. However, we begin with some special cases in which
the operators $O_i(t,s,z^*,w^*)$ turn out to be independent of the
noises $z_t^*,w_t^*$. In such a case, we may write ${\bar
O}_i(t,z^*,w^*) = {\bar O}_i(t)$ and (\ref{generalmaster}) is
indeed a convolutionless, closed master equation
\begin{equation}\label{evorho3}
\partial_t \rho_t = - i[H_{\rm sys},\rho_t] +
[L,\rho_t {\bar O}_1^\dagger(t)] + [{\bar
O}_1(t)\rho_t,L^\dagger]+[L^\dag,\rho_t {\bar O}_2^\dagger(t)] +
[{\bar O}_2(t)\rho_t,L]
\end{equation}
for the non-Markovian open system.

As a direct application of (\ref{evorho3}), let's discuss an
important dephasing process: the system Hamiltonian commutes with
the Lindblad operator $[H_{\rm sys}, \;L]=0$. More generally, we
may consider a solvable model with $[L, H_{\rm sys}]=i\kappa I,
L=L^\dag, \kappa=$ real constant, and $I$ is the identity
operator. For this case, it is easy to check that the solutions to
(\ref{O11}) and (\ref{O12}) are given by \beqa \label{spe1}
O_1(t,s, z^*,w^*)&=&L-\kappa(t-s),\\
O_2(t,s, z^*,w^*)&=&L-\kappa(t-s). \label{spe2}\eeqa By simply
inserting (\ref{spe1}) and (\ref{spe2}) into (\ref{evorho3}), a
closed convolutionless master equation is immediately obtained:
\beq\label{derivation} \frac{d}{dt}\rho_t=-i[H_{\rm sys}, \rh_t]+
f(t)[L\rh_t,L]+f^*(t)[L, \rh_t L]+ g(t)[\rh_t, L] + g^*(t)[L,
\rh_t], \eeq where the time-dependent coefficients are given by
 \beq f(t)=\int^t_0 \al(t-s)ds\;\; {\rm and}\;\;
g(t)=\kappa \int^t_0 \al(t-s)(t-s)ds \eeq Note here that the
finite temperature correlation function $\al(t-s)$ is given by
\beq \al(t-s)=\al_1(t-s)+\al_2(t-s)=\eta(t-s)+i\nu(t-s), \eeq
where \beqa \eta(t-s)&=&\sum_\lam
|g_\lam|^2\coth\left(\frac{\omega_\lam}{2k_BT}\right)\cos[\om_\lam(t-s)],\\
\nu(t-s)&=&-\sum_\lam |g_\lam|^2\sin[\om_\lam(t-s)]. \eeqa

The master equation (\ref{derivation}) can be immediately applied
to qubits decoherence and disentanglement \cite{YE}. To be
specific, let's consider the dephasing process in a quantum
register containing $N$ non-interacting qubits. It is known that
one of the most important decoherence processes in the quantum
registers is that $N$ qubits collectively interact with some
environmental noise which only randomly interrupts the phase of
each qubit. This problem can be effectively modeled by choosing
$H_{\rm sys}=\sum_{i=1}^N \om_i \sigma^z_i$ and $L=\sum_{i=1}^N
\sigma^z_i$ in the general Hamiltonian (\ref{totalH}), where
$\sigma_i^z$ denotes the Pauli matrix for the $i$th qubit.  The
master equation for such a dephasing model is given by
(\ref{derivation}) with $\kappa=0$.

Of particular interest is the Markov approximation where
$\alpha_1(t-s)=\gamma_1 \delta(t-s),
\alpha_2(t-s)=\gamma_2\delta(t-s)$, then
$\bar{O}_1=\frac{\gamma_1}{2}L,
\bar{O}_2=\frac{\gamma_2}{2}L^\dag$, so the resulting master
equation takes the well known Lindblad form \beqa\label{evorho4}
\partial_t \rho_t = -i[H_{\rm sys},\rho_t] &+&\frac{\gamma_1}{2}\left(
2L\rh_tL^\dag -L^\dag L \rh_t- \rh_t L^\dag L\right)\nonumber\\
&+&\frac{\ga_2}{2}\left( 2L^\dag \rh_t L  - L L^\dag \rh_t - \rh_t
L L^\dag \right). \eeqa

Before ending this section we have two remarks in order. First,
let's note that the non-Markovian property of a convolutionless
master equation (\ref{evorho3}) or more generally
(\ref{generalmaster}) is characterized by the time-dependent
coefficients. Second, if the O-operators appearing in (\ref{op})
and (\ref{op1}) or in (\ref{generalmaster}) do contain noises, as
seen from the next section, things become much more complicated.
There is no general guideline how to derive the closed
convolutionless master equation from (\ref{generalmaster}).
However, it has been shown that the evolution equations for the
O-operators $O_i(t,s,z^*,w^*)$ with respect to $s$ rather than $t$
are essential for such a derivation. Here we believe that ``The
existence of a set of uncoupled evolution equations for the
O-operators with respect to $s$" is a very powerful ansatz.
Obviously, this ansatz remains to be tested in more examples
\cite{CRE,SY}.
\section{The damped harmonic oscillator}
\label{dmaped}
\subsection{Stochastic Schr\"odinger equation for the damped harmonic oscillator}
In this section we show that the non-Markovian quantum QSD
approach is versatile enough to handle the general finite
temperature cases with $L\neq L^\dag$. To simplify the
calculations as much as possible, we consider here the simplest
model of this kind  but still worth studying -- the damped
harmonic oscillator model. The model consists of a harmonic
oscillator with frequency $\Omega$, coupled linearly to a large
number of harmonic oscillators: $H_{\rm sys}=\Omega a^\dag a, \,\,
L=a$. We assume that the heat bath is at finite temperature $T$.
This model has been discussed in various contexts (see, e.g.,
\cite{COH,CAR}). In what follows, we use this model to show
explicitly how the finite temperature master equation can be
derived directly from the non-Markovian quantum trajectory
equation. For the damped harmonic oscillator at zero temperature,
it has been shown that $w^*_t=0$ and $\bar{O}_1=F(t)a$,
$\bar{O}_2=0$ \cite{DGS,SY}. Thus the zero-temperature
non-Markovian master equation is immediately obtained from
(\ref{evorho3}). For the finite temperature heat bath, as to be
seen below, the O-operators $O_1$ and $O_2$ for the damped
harmonic oscillator will contain noises $z^*_t$ and $w^*_t$, hence
things become much more involved. We emphasize that, for a quantum
system with $L\neq L^\dag$, the dependence of the O-operators on
the noises is a generic feature for the finite temperature cases.
First, note that the exact quantum state diffusion equation for
the damped harmonic oscillator can be written as \beqa
\label{modelqsd}
\partial_t \psi_t &=& -i\Omega a^\dag a\psi_t +
az^*_t \psi_t -a^\dag \bar{O}_1(t,z^*,w^*)\psi_t\nonumber\\
&&+a^\dag w^*_t \psi_t -a \bar{O}_2(t,z^*,w^*)\psi_t, \eeqa where
$\bar{O}_i$ are defined in (\ref{obar12}).

Crucial to the solution of QSD equation (\ref{modelqsd}) is the
explicit determination of the O-operators. It is easy to check
that the following ansatz will give rise to the solutions of
equation (\ref{O11}) and (\ref{O12}), \beqa \label{oans1}
O_1(t,s,z^*,w^*)&=&f_1(t,s)a+\int^t_0 ds' j_1(t,s,s')
w^*_{s'},\\
O_2(t,s,z^*,w^*)&=&f_2(t,s)a^\dag + \int^t_0 ds' j_2(t,s,s')
z^*_{s'},\label{oans2} \eeqa where $f_i(t,s)$ and $j_i(t,s,s')$
satisfy the {\it coupled nonlinear} integro-differential
equations:
\begin{eqnarray}\label{functions10}
&&\partial_t f_1(t,s) - i\Omega f_1(t,s)
 -f_1(t,s)\int^t_0ds'\al_1(t-s')f_1(t,s')\nonumber \\&-&f_1(t,s)\int^t_0ds'\al_2(t-s')f_2(t,s')
 +\int_0^t ds'\;\alpha_2(t'-s)j_2(t,s',s)= 0, \\
&&\partial_tf_2(t,s) + i\Omega f_2(t,s)
 +f_2(t,s)\int_0^t ds'\;\alpha_1(t-s')f_1(t,s')\nonumber\\&+& f_2(t,s)\int_0^t
 ds'\;\alpha_2(t-s')f_2(t,s')
 +\int^t_0ds''\alpha_1(t-s'')j_1(t,s'',s)=0,\\
&&\partial_t j_1(t,s,s') - f_1(t,s)\int_0^tds''\;\alpha_1(t-s'')j_1(t,s'',s')=0,\\
&&\partial_t j_2(t,s,s') + f_2(t,s)\int_0^t
ds''\;\alpha_2(t-s'')j_2(t,s'',s')=0, \label{functions40}
\end{eqnarray}
with the initial values:
\begin{eqnarray}\label{finals}
&&f_1(t=s,s) = 1,\;\; f_2(t=s,s) = 1,
\\
&&j _1(t=s,s,s') = 0, \;\; j_2(t=s,s,s')=0,\\
&&j_1(t,s,t)
= -f_1(t,s), \;\; j_2(t,s,t)=f_2(t,s).
\end{eqnarray}
Indeed we have seen from (\ref{oans1}) and (\ref{oans2}) that, in
the case of finite temperature,  the O-operators generally involve
the noises, even though at zero-temperature they do not.

With the above evolution equations for $f_i(t,s)$ and $
j_i(t,s,s')$, we are able to simulate the system dynamics by
solving the non-Markovian QSD equation (\ref{modelqsd})
\footnote{More accurately, the numerical simulations need the
nonlinear version of the non-Markovian QSD equation which can be
read off directly from (\ref{modelqsd})}. But our aim in this
paper is to take the quantum trajectories as theoretical tools
rather than as numerical applications. Clearly, the solutions of
the above {\it nonlinear coupled} equations are hard to deal with.
In the next subsection, we will show remarkably that the evolution
equations for $f_i(t,s)$ and  $h_i(t,s,s')$ with respect to $s$
can be turned into a set of first-order linear uncoupled
equations.

\subsection{Decoupling evolution equations for O-operators}

The functions appearing in equations
(\ref{functions10})-(\ref{functions40}), as they stand, are very
difficult to handle analytically.  However, as shown in \cite{SY}
(also see \cite{CRE}), great simplification may arise through
investigating the dependence of the functions on $s$ rather than
$t$. It is proved here that the evolution equations for the
O-operators with respect to $s$ form an uncoupled set of linear
differential equations. The key observation comes from the
Heisenberg operator approach to the non-Markovian QSD
\cite{SY,CRE}. For our purpose in this paper, the main task here
is to find the evolution equations for
\begin{equation}\label{Aop2}
A(s) = \langle z|\langle w|{\mathcal U}_t a(s)|0\rangle =
O_1(t,s,z^*,w^*) \langle z|\langle w |{\mathcal U}_t|0\rangle,
\end{equation}
and
\begin{equation}\label{Aop3}
B(s) = \langle z|\langle w|{\mathcal U}_t a^\dag(s)|0\rangle =
O_2(t,s,z^*,w^*) \langle z|\langle w|{\mathcal U}_t|0\rangle,
\end{equation}
where ${\mathcal U}_t$ is the unitary operator for the extended
system (\ref{totalH2}): $|\psi_{\rm tot}(t)\ra={\mathcal
U}_t|\psi_{\rm tot}(0)\ra$. Clearly, the operators $A(s)$ and
$B(s)$ depend on the time $t$, but for our purpose, we regard
them, and therefore also the operators $O_i(t,s,z^*,w^*)$, as
functions of $s$. The time $t$ appears as a parameter only.
Obviously, the evolution equations for $A(s)$ and  $B(s)$ are
equivalent to that for $O_1(t,s,z^*,w^*)$ and $O_2(t,s,z^*,w^*)$
with respect to $s$, respectively. First, note that the Heisenberg
equation of motion for $a(s)$ gives rise to
\begin{equation}\label{secondorder}
i\partial_s a(s) = \Omega a(s) - \sum_\lambda\sqrt{\bar{n}_\lam
+1}\; g_\lambda e^{-i\omega_\lambda s}d_\lam(s)+ \sum_\lambda
\sqrt{\bar{n}_\lam}\;g_\lambda e^{-i\omega_\lambda
s}e^\dag_\lam(s).
\end{equation}
Therefore, we get \beqa \label{firstorder} i\partial_s A(s) =
\Omega A(s) &+&\sum_\lambda\sqrt{\bar{n}_\lam +1}\; g_\lambda
e^{-i\omega_\lambda s}\langle z|\langle w|{\mathcal
U}_td_\lam(s)|0\rangle\nonumber \\
&+& \sum_\lambda \sqrt{\bar{n}_\lam}\;g_\lambda
e^{-i\omega_\lambda s}\langle z|\langle w|{\mathcal U}_t
e^\dag_\lam(s)|0\rangle. \eeqa Simply integrating the Heisenberg
equation of motion for the environmental annihilation operator
$d_\lambda(s)$, we get
\begin{equation}\label{ingred1}
\langle z|\langle w| {\mathcal U}_t d_\lambda(s)|0\rangle =
-ig^*_\lambda\sqrt{\bar{n}_\lam+1} \int_0^s
ds'\;e^{i\omega_\lambda s'} \langle z|\langle w| {\mathcal U}_t
a(s')|0\rangle,
\end{equation}
where we have used the fact that initially $d_\lambda(0)|0\rangle
= d_\lambda|0\rangle = 0$.

Next, we note that the operator $e_\lambda^\dagger(s)$ in
(\ref{firstorder}) can be dealt with in a similar fashion. In
order to find a closed equation, however, we have to carry out an
expansion at the {\it final} time $t$ rather than the initial
value $s=0$. This simplifies the calculations: $\langle z|\langle
w| {\mathcal U}_t e_\lambda^\dagger(t)|0\rangle =w^*_\lambda
\langle z|\langle w|{\mathcal U}_t|0\rangle$. Thus, we integrate
the Heisenberg equation of motion for $e_\lambda^\dagger(s)$ given
the {\it final} value at $s=t$ to get \beq e_\lambda^\dagger(s) =
e_\lambda^\dagger(t) -i\sqrt{\bar{n}_\lam}g^*_\lambda\int_s^t
ds'\;e^{-i\omega_\lambda s'} a(s'),\eeq hence, we get
\begin{equation}\label{ingred2} \langle z|\langle w| {\mathcal U}_t
e_\lambda^\dagger(s)|0\rangle = w_\lambda^*\langle z|\langle w|
{\mathcal U}_t|0\rangle -{i}\sqrt{\bar{n}_\lam
}\;g_\lambda\int_s^t ds'\;e^{-i\omega_\lambda s'} \langle
z|\langle w| {\mathcal U}_t a(s')|0\rangle.
\end{equation}

Combining the results (\ref{firstorder}), (\ref{ingred1}), and
(\ref{ingred2}), we obtain a linear first-order differential
equation for $A(s)$:
\begin{eqnarray}\label{firstQBMo}
& & \partial A(s) + \Omega A(s) \\ \nonumber & & -i\int_0^s ds' \;
\alpha_1(s-s') A(s') -i\int_s^t ds' \; \alpha_2(s'-s) A(s') = -i
w_s^* \langle z|\langle w| {\mathcal U}_t|0\rangle,
\end{eqnarray}
for $s\in [0,t]$ and a fixed final time $t$. Note that equation
(\ref{firstQBMo}) has to be solved with the final value $A(s=t) =
a \langle z|\langle w|{\mathcal U}_t|0\rangle$.

By noting that $A(s) = O_1(t,s,z^*,z^*)\langle z|\langle
w|{\mathcal U}_t|0\rangle$ from (\ref{Aop2}), so with the derived
first order differential equation for $A(s)$, we may immediately
obtain the evolution equation with respect to $s$ for the desired
operator $O_1(t,s,z^*,w^*)$, we find
\begin{eqnarray}\label{o1operaor}
 i\partial_s O_1(t,s,z^*,w^*)&=& \Omega O_1(t,s,z^*,w^*)-
 i\int_0^s ds' \; \alpha_1(s-s') O_1(t,s',z^*,w^*)\nonumber \\
 && - i\int_s^t ds' \;
\alpha_2(s'-s) O_1(t,s',z^*,w^*) + i w_s^*,
\end{eqnarray}
with the final value
\begin{equation}\label{ofinal1}
O_1(t,s=t,z^*,w^*) =a.
\end{equation}
Similarly, we get the evolution equation for $O_2(t,s,z^*,w^*)$:
\begin{eqnarray}\label{o2operator}
 i\partial_s O_2(t,s,z^*,w^*)&=& -\Omega O_2(t,s,z^*,w^*)+
 i\int_0^s ds' \; \alpha_2(s-s') O_2(t,s',z^*,w^*)\nonumber\\
 && + i\int_s^t ds' \;
\alpha_1(s'-s) O_2(t,s',z^*,w^*) -i z_s^*,
\end{eqnarray}
with the final value
\begin{equation}\label{ofinal2}
O_2(t,s=t,z^*,w^*) =a^\dag.
\end{equation}

Remarkably, we see that the coupled nonlinear equations
(\ref{O11}) and (\ref{O12}) have become a set of uncoupled linear
equations (\ref{o1operaor}) and (\ref{o2operator}). We emphasize
that equations (\ref{o1operaor}) and (\ref{o2operator}), with the
final values  (\ref{ofinal1}) and (\ref{ofinal2}) respectively,
are all that is required for the derivation of the master equation
for the damped harmonic oscillator.

The equations (\ref{o1operaor}) and (\ref{o2operator}) in terms of
$f_1(t,s), f_2(t,s)$ and $h_1(t,s,s'), h_2(t,s,s')$ can be written
as a set of uncoupled {\it linear} equations for $s\in[0,t]$, with
$t$ fixed,
\begin{eqnarray}\label{functions_s}
\partial_s f_1(t,s) + i\Omega f_1(t,s)
 + \int_0^s ds'\;\alpha_1(s-s')f_1(t,s')
 + \int_s^t ds'\;\alpha_2(s'-s)f_1(t,s')&  = & 0, \\
\partial_s f_2(t,s) -i\Omega f_2(t,s)
 -\int_0^s ds'\;\alpha_2(s-s')f_2(t,s')
 - \int_s^t ds'\;\alpha_1(s'-s)f_2(t,s')&  = & 0, \\
\partial_s j_1(t,s,s') +i \Omega j_1(t,s,s')
 +\int_0^s ds''\;\alpha_1(s-s'')j_1(t,s'',s')&&\\ \nonumber
 +\int_s^t ds''\;\alpha_2(s''-s)j_1(t,s'',s') = \delta(s-s'),\\
\partial_s j_2(t,s,s') -i \Omega j_2(t,s,s')
 - \int_0^s ds''\;\alpha_2(s-s'')j_2(t,s'',s')&&\\ \nonumber
 - \int_s^t ds''\;\alpha_1(s''-s)j_2(t,s'',s') = -\delta(s-s'),
\end{eqnarray}
with the final values
\begin{eqnarray}\label{finals12}
& &f_1(t,s=t) = 1,\;\;  f_2(t,s=t) = 1,\\
& & j_1(t,t,s)= 0,\;\; j_2(t,t,s) = 0,\;\;\; \mbox{for all}\;\; s.
\end{eqnarray}

\subsection{Master equation for the damped harmonic oscillator}

Now we are going to derive the exact master equation for the
damped harmonic oscillator. We will see that the evolution
equations of $O_i$ are crucial for deriving the corresponding
non-Markovian master equation from the stochastic schrodinger
equation (\ref{modelqsd}).  From (\ref{generalmaster}), we get
\beqa\label{generalmaster10}
\partial_t \rho_t &=& -i\Omega[a^\dag a,\rho_t]
+[a,\,\, {\mathcal M}\{P_t {\bar O_1}^\dagger(t, z^*,w^*)\}]
-[a^\dagger,\,\,{\mathcal M}\{{\bar
O}_1(t,z^*,w^*) P_t\}]\nonumber\\
&&+[a^\dag,\,\,{\mathcal M}\{P_t{\bar O}_2^\dag(t,z^*,w^*)\}] -
[a,\,\,{\mathcal M}\{\bar{O}_2(t,z^*,w^*)P_t\}]. \eeqa

For the equation (\ref{generalmaster10}) to be a closed,
convolutionless master equation, we need the explicit expressions
for the ensemble means:
\begin{equation}\label{ensmean}
R_1(t,s)\equiv {\mathcal
M}[O_1(t,s,z^*,w^*)P_t],\;\;R_2(t,s)\equiv {\mathcal
M}[O_2(t,s,z^*,w^*)P_t].
\end{equation}
For this purpose, we take the mean ${\mathcal M}[\cdots P_t]$ of
equation (\ref{o1operaor}), and then we get
\begin{eqnarray}\label{Req1}
&&\partial_s R_1(t,s) + i\Omega R_1(t,s) \\ \nonumber &&+i\int_0^s
ds'\;\alpha_1(s-s')R_1(t,s') -i\int_s^t ds'\;\alpha_2(s'-s)
R_1(t,s') = -i{\mathcal M}[w^*_s P_t].
\end{eqnarray}
Thus, we have to find an expression for ${\mathcal M}[z^*_s P_t]$.
By using Novikov's theorem \cite{NOV}, it is possible to relate it
to the $O$-operator again \cite{YDGS}:
\begin{equation}\label{meanzP}
{\mathcal M}[w^*_s P_t] = \int_0^t ds' \alpha_2^*(s-s') {\mathcal
M}[P_t O_2^\dagger(t,s',z^*,w^*)]
 = \int_0^t ds' \alpha_2^*(s-s') R_2^\dagger(t,s').
\end{equation}
Thus, we find the equations for the operator $R_1(t,s)$, as
function of $s$, to be
\begin{eqnarray}\label{Req10}
&&\partial_s R_1(t,s) + i\Omega R_1(t,s) \\ \nonumber &&+\int_0^s
ds'\;\alpha_1(s-s')R_1(t,s') +\int_s^t ds'\;\alpha_2(s'-s)
R_1(t,s') = \int^t_0ds'\alpha^*_2(t-s')R_2^\dag(t,s').
\end{eqnarray}
and similarly, for the operator $R_2(t,s)$, we have
\begin{eqnarray}\label{Req20}
\partial_s R_2(t,s)&-& i\Omega R_2(t,s)-\int_0^s
ds'\;\alpha_2(s-s')R_2(t,s')\\
&-&\int_s^t ds'\;\alpha_1(s'-s) R_2(t,s') =
 -\int_0^t ds' \alpha_1^*(s-s') R_1^\dagger(t,s'). \nonumber
\end{eqnarray}
Note that the equations (\ref{Req10}) and (\ref{Req20}) have to be
solved with ``final values''
\begin{equation}\label{Rfinals}
R_1(t,s)|_{s=t} = a\rho_t,\;\;R_2(t,s)|_{s=t} = a^\dag\rho_t,
\end{equation}
as one can easily see from (\ref{ofinal1}) and (\ref{ofinal2}).
Note that equations  (\ref{Req10}) and (\ref{Req20}) are a set of
coupled linear equations that are expected from (\ref{oans1}) and
(\ref{oans2}).

By observing (\ref{oans1}) and (\ref{oans2}), we find that the
solutions of the crucial equations (\ref{Req10}) and (\ref{Req20})
with the final values (\ref{Rfinals}) must take the following
form:
\beqa\label{Rexp} R_1(t,s) &=& F(t,s)a\rho_t + G(t,s) \rho_ta,\\
R_2(t,s) &=& H(t,s)a^\dag\rho_t + I(t,s) \rho_t a^\dag,
\label{Rexp1} \eeqa where $F(t,s), G(t,s), H(t,s),$ and $I(t,s)$
are complex functions with the
appropriate final values at $s=t$: \beqa F(t,s=t)&=&1,\,\,\,G(t,s=t)=0,\\
H(t,s=t)&=&1,\,\,\, I(t,s=t)=0.
 \eeqa

Once we have $R_i(t,s)={\mathcal
M}[O_i(t,s,z^*,w^*)P_t]\,\,(i=1,2))$ in the form of (\ref{Rexp})
and (\ref{Rexp1}), it is straightforward to write down the master
equation. Simply inserting result (\ref{Rexp}) and (\ref{Rexp1})
into the general form of the non-Markovian QSD-master equation
(\ref{generalmaster10}), we get \beq\label{generalmaster20}
\partial_t \rho_t = -i\Omega[a^\dag a,\rho_t]
+a(t)[a,\,\, \rh a^\dag] +b(t)[a,\,\,a^\dag\rh]
+c(t)[a^\dag,\,\,a\rh] +d(t)[a^\dag,\,\, \rh a], \eeq with
\begin{eqnarray}\label{qbmfunctions}
a(t) & = & \int_0^t ds \Big(\alpha_1^*(t-s)F^*(t,s) -
\alpha_2(t-s)I(t,s)\Big),
\\
b(t) & = & \int_0^t ds
\Big(\alpha^*_1(t-s)G^*(t,s)-\alpha_2(t-s)H(t,s)\Big),
\\
c(t) & = & -a^*(t),
\\
d(t) & = & -b^*(t).
\end{eqnarray}
The equation (\ref{generalmaster20}) is valid for arbitrary
temperature, so it is expected to provide a good description for
the low temperature regimes where the Markov approximation is
doomed to fail. Clearly, the master equation
(\ref{generalmaster20}) derived in this way automatically
preserves the positivity, Hermiticity and trace. Moreover, it can
be easily shown that the above master equation is of the Lindblad
form, but the coefficients are time-dependent, so it depicts the
non-Markovian dynamics.

\subsection{Determination of the coefficients of master equation}
In what follows, we show that the coefficients $a(t),b(t),c(t)$
and $d(t)$ can be expressed in terms of solutions of some basic
equations.  Suppose that $u(t,s)$ is the solution to the following
homogeneous equation: \beq\label{beqn}
\partial_s u(t,s) -i \Omega u(t,s)+ \int^t_s ds'\beta(s,s')u(t,s')=0, \eeq
with the final value $u(t,s=t)=1$. Here the kernel function
$\beta(s,s')$ is defined as \beq
\beta(s,s')=\alpha_2(s-s')-\alpha_1(s'-s). \eeq Then from
(\ref{Req10}) and (\ref{Req20}) it can be shown that $F(t,s),
G(t,s), H(t,s)$ and $I(t,s)$ satisfy the following inhomogeneous
equation, \beqa \label{eq10} \partial_s F(t,s) + i\Omega
F(t,s)-\int_0^s
ds'\beta(s',s)F(t,s')&=&-X(t,s),\\
\label{eq20} \partial_s G(t,s) + i\Omega G(t,s)-\int_0^s
ds'\beta(s',s)G(t,s')&=&X(t,s),\\
\partial_s H(t,s) - i\Omega
H(t,s) -\int_0^s
ds'\beta(s,s')H(t,s')&=&Y(t,s),\label{eq30}\\
\partial_s I(t,s)-i\Omega I(t,s)-\int_0^s
ds'\beta(s,s')I(t,s')&=&-Y(t,s),
 \label{eq40} \eeqa
where the functions $X(t,s)$ and $Y(t,s)$ are given by \beqa
X(t,s)&=&\int_0^tds'\alpha_2(s'-s)u^*(t,s'),\\
Y(t,s)&=&\int^t_0ds'\alpha_1(s'-s)u(t,s').\eeqa Hence it is easy
to see that the coefficients of the master equation
(\ref{generalmaster20}) can be expressed in terms of the solutions
of those basic equations (\ref{beqn})  and
(\ref{eq10})-(\ref{eq40}).


\section{Conclusions}\label{sect5}

The non-Markovian quantum trajectories offer a brand new method
for exploring a quantum system coupled to a non-Markovian
environment. Such a situation appears in various physical
problems, such as materials with photonic band gaps or output of
atom lasers. It is known that master equations and quantum
trajectories in Markov regimes are of fundamental importance for
the description of open system dynamics. Moreover, they are often
complementary to each other to provide a full picture of the
underlying physics. In this paper we show that in the case of
general finite temperature heat bath this fruitful interrelation
between the master equations and the quantum trajectories can also
be established in the non-Markovian regimes. In particular, we
show that by averaging the stochastic Schr\"odinger equation the
exact non-Markovian master equation of the damped harmonic
oscillator at finite temperature can be obtained. The master
equation derived in this way takes the convolutionless form, so
the non-Markovian property is encoded in the time-dependent
coefficients.

In the Markov regimes, the derivation of a master equation poses
no special difficulties. However, it is a rather difficult task to
derive a convolutionless exact master equation in the
non-Markovian regimes. As shown in this paper, non-Markovian
quantum trajectories provide a very useful tool in handling exact
or approximate non-Markovian master equations. We believe that the
techniques employed in this paper may find broader applications to
other quantum open systems, leading to the establishment of a
general master equation for the system interacting with a bosonic
heat bath. This will be the topic of future investigations.

\section*{Acknowledgments}
I thank Lajos Di{\'o}si, Ian Percival and Walter Strunz for
valuable discussions and I particularly thank Joseph Eberly for
reading the manuscript and for his critical comments. Grants from
the NEC Research Institute (US) and Leverhulme Foundation (UK)
provided financial support.


\begin{thebibliography}{99}

\bibitem{GAR} C.\ Gardiner and P.\ Zoller,
                   {\it Quantum Noise}, second edition, (Springer, Berlin) (2000).

\bibitem{CAR1} H.\ Carmichael,
                   {\it Statistical Methods in Quantum Optics  I}, (Springer, Berlin) (2000).

\bibitem{COH} C.\ Cohen-Tannoudji, J.\  Dupont-Roc, G.\ Grynberg, {\it Atom and Photon Interactions:
Basic Processes and Applications},  (John Wiley \& Sons, Inc. New
York ) (1998).

\bibitem{LIN} G.\ Lindblad,
                   Comm. Math. Phys. {\bf 48}, 119 (1976).

\bibitem{GKS}
V. Gorini, A. Kossakowski, and E. C. G. Sudarshan, J. Math. Phys.
{\bf 17}, 821 (1976).

\bibitem{CAR} H.\ Carmichael, {\it An Open System Approach to
                   Quantum Optics}, (Springer, Berlin) (1993).


\bibitem{DM}J.\ Dalibard, Y.\  Castin, and K.\ M{\o}lmer,
                  Phys. Rev. Lett. {\bf 68} 580 (1992).

\bibitem{GP}N.\ Gisin and I.\ C.\  Percival,
                  J. Phys. A {\bf 25}, 5677 (1992); {\bf 26}, 2233 (1993).

\bibitem{GZ}C.\ Gardiner, A.\ Parkins, and P.\ Zoller,
              Phys. Rev. A {\bf 46}, 4363 (1992).

\bibitem{WM}H.\ M.\ Wiseman and G.\ J.\ Milburn, Phys. Rev. A {\bf 47},
1652 (1993).


\bibitem{PK} M.\ Plenio and P.\ L.\ Knight,
                   Rev. Mod. Phys. {\bf 70}, 101 (1998).

\bibitem{PER} I.\ C.\  Percival, {\it Quantum State Diffusion},
(Cambridge, England) (1998).


\bibitem{NM1} S.\ John and T.\ Quang, Phys. Rev. Lett. {\bf 74}, 3419; B.\  M.\ Garraway,
Phys. Rev. A {\bf 55}, 4636 (1997); M.\ Nikolopoulos and P.\
Lambroupoulos, Phys. Rev. A  {\bf 61}, 053812 (2000).

\bibitem{NM2}J.\ J.\ Hope, Phys. Rev. A {\bf 55}, R2531 (1997);
H.-P.\ Breuer,\ D.\ Faller, B.\ Kappeler, and F.\ Petruccione,
Phys.\ Rev.\ A  {\bf 60}, 3188 (1999).

\bibitem{NAK} S. Nakajima, Prog. Theor. Phys. {\bf 20}, 948 (1958).

\bibitem{ZWA} R. Zwanzig, J. Chem. Phys. {\bf 33}, 1338 (1960).


\bibitem{DS}   L.\ Di\'osi and W.\ T.\ Strunz,
                  Phys. Lett. A {\bf 235}, 569 (1997).

\bibitem{DGS}  L.\ Di\'osi, N.\  Gisin  and W.\  T.\ Strunz,
                 Phys. Rev. A {\bf 58}, 1699 (1998).

\bibitem{DGS1}  W.\ T.\ Strunz, L.\ Di\'osi, and N.\ Gisin,
                  Phys. Rev. Lett. {\bf 82}, 1801 (1999).

\bibitem{YDGS}  T.\ Yu,  L.\ Di\'osi, N.\  Gisin  and W.\  T.\ Strunz,
                  Phys. Rev. A  {\bf 60}, 91 (1999).

\bibitem{YDGS2}  W.\ T.\  Strunz, L.\ Di\'osi, N.\  Gisin  and T.\ Yu,
                Phys. Rev. Lett.  {\bf 83}, 4909 (1999).

\bibitem{BKP}   H.\ Breuer,  B.\ Kappler, and F.\ Petruccione,
                  Phys. Rev. A  {\bf 59}, 1633 (1999).

\bibitem{JCW} M.\ Jack, M.\ Collet, and D.\  Walls.
                    Phys. Rev. A {\bf 59}, 2308 (1999).

\bibitem{CRE}  J.\ Cresser,
                  Laser Phys.  {\bf 10}, 1 (2000).

\bibitem{BU}   A.\  A.\  Budini, Phys. Rev.  A {\bf 63}, 012106
(2000).

\bibitem{GW}   J.\  Gambetta and H.\ M.\ Wiseman, Phys. Rev.  A {\bf 66}, 012108
(2002); {\bf 66}, 052105 (2002); {\bf 68}, 062104 (2003).

\bibitem{BA}   A.\ Bassi, Phys. Rev.  A {\bf 67}, 062101
(2003).

\bibitem{YDGS1}  T.\ Yu, L.\ Di\'osi, N.\  Gisin  and W.\ T.\ Strunz, Phys. Lett.  A  {\bf 265}, 331 (2000).

\bibitem{SY}  W.\ T.\ Strunz and T.\ Yu, Preprint, arXiv: quant-ph/0312103 (2003),
Physical Review A (to appear).


\bibitem{HR}For the quantum Brownian motion model, a collection of references are as follows:
            R. Feynman and F.\ L.\ Vernon, Ann.  Phys. (N.Y.) {\bf 24}, 118
            (1963);
            A.\ O.\ Caldeira and A.\ J.\ Leggett, Physica A {\bf 121}, 587 (1983);
            H.\ Grabert, P.\ Schramm, and
            G.-L.\ Ingold, Phys. Rep. {\bf 168}, 115 (1988);
            F.\ Haake and R.\ Reibold,
            Phys. Rev. A {\bf 32}, 2462 (1985); W.\ G.\ Unruh and W.\ Zurek,
            Phys. Rev. D {\bf 40}, 1071 (1989); B.\ L.\ Hu, J.\ P.\ Paz and Y.\ Zhang,
            Phys. Rev. D {\bf 45}, 2843
            (1992); Phys. Rev. D {\bf 47}, 1576 (1993); J. J. \ Halliwell and T. Yu,
            Phys. Rev. D {\bf 53}, 2012 (1996); G.\ W.\ Ford and R.\ F.\ O'Connell,
            Phys. Rev.   D  {\bf 64},  105020 (2001).

\bibitem{KK}G.\ W.\ Semenoff and H.\ Umezawa,  Nucl.  Phys.  B
{\bf 220}, 196 (1983).

\bibitem{YE} T.\ Yu and J.\ H.\ Eberly,
                Phys. Rev.  B  {\bf 66}, 193306 (2002).

\bibitem{NOV}A.\ Novikov, Sov. Phys.  JETP {\bf 20}, 1290 (1965).

\end{thebibliography}
\end{document}